\documentclass[aps,prb,twocolumn,showpacs,superscriptaddress,floatfix]{revtex4}

\usepackage{graphicx}
\usepackage{amssymb}
\usepackage{amsmath}
\usepackage{amsfonts}
\usepackage{amsthm}
\usepackage{verbatim}
\usepackage{bbm}
\usepackage{subfigure}

\newfont{\BBB}{msbm10 scaled\magstephalf}

\newcommand{\Z}{\mathbb{Z}}

\newcommand{\C}{\mathbb{C}}
\newcommand{\tens}{\times}

\newcommand{\te}{\theta}
\newcommand{\sd}{\rtimes}
\newcommand{\Rcal}{\mathcal{T}_r}
\newcommand{\Tcal}{\mathcal{T}}
\newcommand{\Tright}{\mathcal{T}_r}
\newcommand{\Acal}{\mathcal{A}}
\newcommand{\Ucal}{\mathcal{U}}

\newcommand{\unitmatrix}{\textbf{1}}
\newcommand{\Od}{\tilde{O}}
\newcommand{\Id}{\tilde{I}}
\newcommand{\Td}{\tilde{T}}
\newcommand{\Hd}{\tilde{H}}
\newcommand{\Gd}{\tilde{G}}
\newcommand{\Orb}{\mathcal{O}}
\newcommand{\braid}{\mathcal{R}}
\newcommand{\eps}{\varepsilon}

\begin{document}
%\title{{\normalsize {\hfill ITFA-2005-44}}\\[2.cm]{\Large\bf Nematic phases and the breaking of double symmetries}}
\title{Nematic phases and the breaking of double symmetries}
\author{C.J.M.~Mathy}
\affiliation{Department of Physics, Princeton University, Jadwin
Hall, Princeton, NJ 08544, United States} 
\date{January 25, 2006}
\author{F.A.~Bais}
\affiliation{ Institute for Theoretical Physics, Valckenierstraat
65, 1018 XE Amsterdam, The Netherlands}

\begin{abstract}
\vspace{0.3in} \noindent In this paper we present a phase
classification of (effectively) two-dimensional non-Abelian
nematics, obtained using the Hopf symmetry breaking formalism. In
this formalism one exploits the underlying double symmetry which
treats both ordinary and topological modes on equal footing, i.e. as
representations of a single (non-Abelian) Hopf symmetry. The method
introduced in the literature \cite{Bais:2002pb,Bais:2002ny} and
further developed in a paper published in parallel
\cite{bmbreaking:2006} allows for a full classification of defect
mediated as well as ordinary symmetry breaking patterns and a
description of the resulting confinement and/or liberation
phenomena. After a summary of the formalism, we determine the double
symmetries for tetrahedral, octahedral and icosahedral nematics and
their representations. Subsequently the breaking patterns which
follow from the formation of admissible defect condensates are
analyzed systematically. This leads to a host of new (quantum and
classical) nematic phases.  Our result consists of a listing of
condensates, with the corresponding intermediate residual symmetry
algebra $\Rcal$ and the symmetry algebra $\mathcal{U}$
characterizing the effective ``low energy'' theory of surviving
unconfined and liberated degrees of freedom in the broken phase. The
results suggest that the formalism is applicable to a wide variety
of two dimensional quantum fluids, crystals and liquid crystals.
\vspace{0.6in}
\end{abstract}
\maketitle \footnotetext[1]{bais@science.uva.nl}
\footnotetext[2]{cmathy@princeton.edu}
\section{Introduction}
The goal of this paper is to apply a Hopf symmetry breaking analysis
to defect condensates in nematics. The subject of quantum liquid
crystal phases is quite extensive, and we first highlight some
nematic phases where our methods are the most relevant. Then we
describe how Hopf symmetries (in particular, double symmetries)
characterize the degrees of freedom in phases with spontaneously
broken symmetry, and discuss the Hopf symmetries relevant to the
exotic nematic phases we focus on in this work. Finally, we describe
the formalism for symmetry breaking, and work out all possible
defect mediated phase transitions in exotic achiral nematics, be
they classical or quantum. Such phases have only been observed in
``classical'' systems, in which we do not expect quantum
superpositions of defects to form condensates. Still, we choose
these phases because they highlight the power and generality of our
approach. There are known examples of quantum liquid crystals in
which a rank two tensor order parameter field is needed to describe
the phase, and our methods are definitely applicable to such
systems, though capturing the exotic nonabelian phases requires at
least a third rank tensor.

\subsection{Defect condensates}

Classical liquid crystals have been studied for a long time.
Recently there has been a renewed interest in exotic liquid
crystals, and they have been invoked to explain certain phases in
bent-core liquid crystals. An exhaustive analysis of ordinary
symmetry breaking patterns in classical nematics can be found in the
literature\cite{Lubensky:bent-core}. In a sense, our study
complements this analysis, as we work out an exhaustive analysis of
defect mediated phase transitions and their interpretation as the
breaking of certain double symmetries.

The literature on liquid crystal phases in quantum Hall systems has
vastly grown in recent years \cite{Kivelson,Radz}. In High-Tc
superconductivity, the stripe phase is a two dimensional quantum
smectic, and recently a theoretical study has analyzed the
possibility of having a topological nematic phase in such a system
\cite{Zaanen}. The nematic order is arrived at by a defect
condensate from a crystalline phase, in other words through a non
trivial vacuum expectation value of some disorder parameter. This in
contrast with the more conventional spin nematic order, known to
exist in superfluids \cite{Coleman} and High-Tc superconductors
\cite{Zaanen2}, where a biaxial nematic phase has been found.
Research on spin nematic order is still vibrant.

Our work offers a general approach to the study of all conceivable
condensate mediated phase transitions using an analysis similar in
vein to Landau theory. Just like Landau theory, it could serve as
the kinematical backbone of more detailed dynamical studies
(involving effective Hamiltonians and a renormalization group
analysis, for example). In this paper we  focus on exotic nematics
whose residual symmetry correspond to the tetrahedral, octahedral
or icosahedral group. We choose such phases to expose the power of
our method, and because as described above there is an ever
growing interest in quantum liquid crystal phases.

\section{Nematic phases and double symmetries}

In this section we define nematic phases quite generally in terms of
symmetries, focusing on systems that are effectively two
dimensional. We then briefly recall how the representation theory of
an underlying Hopf (quantum double) symmetry leads to a description
that treats regular excitations, topological point defects and dyons
on equal footing (for details, see a paper published in
parallel\cite{bmbreaking:2006}). On rather general grounds we
determine the Hopf symmetries that characterize the relevant nematic
phases. The precise outcome for nematics is a \emph{modified quantum
double}, which is a variation on Drinfeld's quantum double of a
group\footnote{We give a tailor made summary of these double
algebras in Appendix A of a related paper\cite{bmbreaking:2006}.}.
The generality of the approach makes our methods applicable to
basically all ``nematics'', be they classical or quantum, global or
gauged. In fact, any phase that results from spontaneous
condensation phenomena can be subjected to a similar analysis. Our
first task is to classify all modes in these media as irreps of a
Hopf symmetry. This classification serves as a crucial ingredient in
the analysis of the next section, where we study phase transitions
induced by the condensation of defects.

\subsection{Nematic liquid crystals}
A nematic liquid crystal is a phase with complete translational
symmetry, and incomplete rotational symmetry
\cite{deGennes,Mermin,Chaikin}. The phase then inherits the name of
the residual rotational group: if the residual rotational group is
the
tetrahedral group, for example, the phase is called a tetrahedral nematic.\\
The residual internal rotational symmetry group $H$ can be any
proper subgroup of $G=SO(3)$. $H$ is the stabilizer of some fixed
tensor in a representation of $G$. This implies that $H$ must be a
closed subgroup of $SO(3)$. The closed subgroups of $SO(3)$ are
well known:
\begin{equation}
G=SO(3)\mapsto H \in \{C_n, D_n, T, O, I, SO(2)\sd\Z_2\},
\end{equation}
where we used the notation employed in the crystallography
literature\cite{Butler}. $C_n=\Z_n$, the abelian cyclic group of
order $n$. $D_n$ is the dihedral group of order $n$, $T$ is the
tetrahedral group, $O$ the octahedral group, and $I$ the
icosahedral group.

If we want to consider inversion symmetries, then we break
$G=O(3)$ to a closed subgroup $H$ of $O(3)$. In the case of an
achiral tetrahedral nematic, $O(3)$ is broken to $T_d$ (we adopt
the crystallographic notation for the symmetries of a tetrahedron
including reflections).

We will call the residual symmetry group $H$ the \emph{electric
group} of the phase, and in general we will denote an electric
group of a phase by $H_{el}$.

\subsection{Excitations}

In general, in a phase where a group $G$ is spontaneously broken to
a subgroup $H$, one distinguishes between three types of modes:
regular excitations in what is often called the ``electric'' sector
, topological defects corresponding to the magnetic sector, and
mixed excitations in the dyonic sector.

\subsubsection{Regular excitations}

Regular excitations (or regular modes) are smooth, low energy
excitations of the basic fields that characterize the system.
Examples are continuity modes (present because of conservation
laws), and Goldstone modes. These modes may be coupled to each other
(such as is the case in classical nematics \cite{Chaikin} ). We will
often refer to these regular modes as \emph{electric modes}.

The regular modes transform under irreps $\Pi_\alpha$ of the
symmetry group $H$. The corresponding states form a vector space on
which the elements of $H$ act as linear transformations. The states
are denoted as $|\xi^j>$, where the $\xi^j$ stand for all the
numbers we need to characterize the state, spatial coordinates and
other quantum numbers. Multi-particle states are described by tensor
products of the elementary representations which are assumed to be
reducible and can be decomposed into irreducible components given by
the appropriate Clebsch-Gordan series:
\begin{equation}\label{eq:fusionrule}
\Pi_\alpha \otimes  \Pi_\beta  = N_{\alpha\beta}^\gamma \Pi_\gamma
\;,
\end{equation}
These rules for combining representations are also called
\emph{fusion rules}. They imply that if we bring particles $1$ and
$2$, in states $|\xi_1^i>$ and $|\xi_2^j>$ respectively, closely
together, and we measure the quantum numbers of the combined system,
that we can get different outcomes, precisely given by the
Clebsch-Gordan coefficients corresponding to the decomposition
(\ref{eq:fusionrule}).

Let us consider one of the irreps $\Pi_\alpha$, and choose a basis
for the corresponding  vector space denoted by
$\{|e_j^{\alpha}>\}$, where j labels the different basis vectors.
We then write for the (matrix) action of $g\in H$ in that basis:
\begin{equation}
\alpha(g)\cdot |e_j^{\alpha}> =\alpha(g)_j^k\cdot |e_k^{\alpha}>.
\end{equation}
The physical requirements on the representations are that they are
\emph{unitary}, because the state should stay normalized under the
action of $G$. However, if $H$ is not simply connected, then in a
quantum nematic the states can transform under \emph{projective}
representations of $H$. This means that the action of $H$ on the
states is not a group homomorphism: the action of  $g_2$ first and
then $g_1$ is not equal to the action of $g_1g_2$. The actions may
only differ by a phase:
\begin{equation}
\alpha(g_1)\alpha(g_2) \Psi = e^{\varphi(g_1,g_2)}\alpha(g_1g_2)
\Psi
\end{equation}
This is allowed because the phase factor disappears when we
calculate (transition) probabilities $|\Psi|^2$. For example, half
integral spin representations transform under a projective
representation of $SO(3)$. As a matter of fact projective
representations of a group $H$ correspond to (faithful)
representations of $\Hd$, defined as the lift of $H$ in the
universal covering group $\Gd$ of $G$. For example, a spin
$\frac{1}{2}$ particle forms a doublet, which is a faithful
representation of $SU(2)$.

\subsubsection{Topological defects}

Topological point defects in two spatial dimensions (or line
defects in three dimensions) correspond to nontrivial
configurations of some order parameter field which are stable for
topological reasons. The point defects are characterized by the
first homotopy group $\pi_1(G/H)$ of the vacuum manifold $G/H$.
The group element that corresponds to a given defect is called its
\emph{topological charge} or \emph{magnetic flux}. In general we
will denote the magnetic group of a phase by $H_m$.

Using a standard theorem from homotopy theory:
\begin{equation}
\pi_1(G/H)\simeq \pi_0(\Hd),
\end{equation}
we find that the point defects are characterized by the zeroth
homotopy group (which studies the connected components) of $\Hd$. In
particular, if $H$ is a discrete group, then $\Pi_1(G/H)=\Hd$. Let
us assume that $H$ is discrete, then we will call $\Pi_1(G/H)$ the
\emph{magnetic group} $H_m$ of the phase. The first observation we
should make about the composition rule for the defect charges is
that it is specified by the structure of the first homotopy group
and corresponds therefore to group multiplication. We have to say
more about this though, because in the cases of interest these
groups are non-Abelian which at first sight gives rise to unwanted
ambiguities in the fusion rules for defects.

Let us now denote the (internal) physical state corresponding to a
defect charge $g$ by a ket $|g>$.  The defect states form a vector
space $V$ spanned by the $|g>$ with $g\in \Hd$: $V=\{\sum_j
\lambda_j |g_j>: \lambda_j \in\C, g_j\in H \}$\footnote{As our
treatment is quantum mechanical, the scalars of the vector space are
in $\C$.}. The vector space spanned by group elements and equipped
with the group multiplication is called the \emph{group algebra} and
denoted $\mathbb{C}H$.

In our quantum treatment it can make sense to add certain defect
states. A state $|g_1>+ |g_2>$, for example, would correspond to a
quantum superposition of the defects $g_1$ and $g_2$. A priori there
is no obvious classical interpretation for this superposition. We
note however that there are actually cases where superpositions of
defects can be given a classical
interpretation\cite{bmmelting:2006}.

States in different irreps cannot form a superposition: the span
of states in one irrep forms a superselection sector. To figure
out how many defects are in one sector, we need to know how the
electric group $H_{el}$ acts on the magnetic group $H_m$. As a
matter of fact, if we have a defect in our system and act on the
system with a global symmetry transformation, then we may obtain a
different defect. Given $h\in H_{el}$ and $g\in H_m$, we denote
the action of $h$ on $g$ by $h\cdot g$. This action satisfies

\begin{eqnarray}
&& h_1 \cdot (h_2 \cdot g) = (h_1h_2)\cdot g \quad \forall h_1, h_2 \in H_{el}, g \in H_m \nonumber\\
&& h\cdot (g_1g_2)=(h\cdot g_1)(h\cdot g_2) \quad \forall h\in
H_{el}, g_1,g_2\in H_m \nonumber
\end{eqnarray}

The first equation is natural, it simply says that the group
$H_{el}$ acts on $H_m$ as a group. The second equation implies
that the action of a global symmetry transformation $h$ on a
configuration that is composed of two defects next to each other,
with topological charges $g_1$ and $g_2$, is equal to the action
on each defect separately with $h$.

For example, if $H_m = H_{el} \equiv H$, the action of a global
symmetry transformation $h\in H$ on defect $|g>$ is a group
conjugation of the topological charge:
\begin{equation}
h\cdot |g> = |hgh^{-1}>
\end{equation}

The defect representations are therefore labelled by the defect
classes $A$ in $H_m$\cite{Bais:1980fm}. These classes correspond to
sets of defects that are transformed into each other under the
action of $H_{el}$. We should think of the classes $A$ as defect
representations $\Pi^A$, and they represent $H_{el}$ invariant
sectors of the theory.

At this stage of the analysis the defect representations $\Pi^A$
might be reducible. However, the algebra can be extended with other
operators in our theory which make the classes irreducible
representations. We can in principle measure the precise flux of a
defect, using (global) Aharonov-Bohm scattering
experiments\cite{dwpb1995,Preskill}. Correspondingly there exist
certain \emph{projection operators} $P_g$ in our theory ($g\in
H_m$), that act on the defects according to
\begin{equation}
P_g |g'> = \delta_{g,g'}|g'>.
\end{equation}
These projection operators span a vector space which is isomorphic
to the vector space of functions from $H_m$ to $\C$, which we denote
by $F(H_m)$. Namely $P_g$ can be associated with the function on
$H_m$ defined by $P_g(g')=\delta_{g,g'}$. $F(H_m)$ can be turned
into an algebra by taking pointwise multiplication, and this is
precisely the algebra of the projection operators! Thus we will
associate the projection operators with $F(H_m)$.

The defect classes form irreps under the combined action of $H_{el}$
and $F(H_m)$. Note that we've described the action of $F(H_m)$ and
$H_{el}$ separately, and we need to know what happens when a
projection operator and a global symmetry transformation are applied
in succession. Thus we want to turn the combination of $H_{el}$ and
$F(H_m)$ into an \emph{algebra}, i.e. we want to be able to multiply
elements of $H_{el}$ and $F(H_m)$. Physics dictates what the answer
is: the multiplication in this algebra is set by \cite{dwpb1995}
\begin{equation}
h P_g = P_{h\cdot g} h.
\end{equation}
The physical motivation for this equation is as follows: if we
measure a flux $g$ with $P_g$, and then conjugate the defect with
$h$, we have a flux $h\cdot g$. This action is equivalent to first
acting on the
defect with $h$, and then measuring $h\cdot g$ with $P_{h\cdot g}$.\\
We call the algebra defined in this way a \emph{modified quantum
double} (because it closely resembles the quantum double $D(H)$,
which is a special case of the modified quantum double with
$H_m=H_{el}=H$), and denote it by\footnote{In the mathematical
literature this product is often written as a ``bow tie'', $F(H_m)
\bowtie \C H_{el}$ indicating that there is a nontrivial action
defined of the factors on each other. Formally the structure
corresponds to an example of a bicrossproduct\cite{Majid} of the
Hopf algebras $F(H_m)$ and $\C H_{el}$. To keep the notation simple,
we use an ordinary product sign, to clearly distinguish it from the
tensor product sign which we will use to describe  multiparticle
states.} $F(H_m) \times \C H_{el}$. As a vector space, the modified
quantum double is simply $F(H_m) \otimes \C H_{el}$. The
multiplication is set by the action defined above. Thus
we conclude that \emph{the defects transform under irreps of $F(H_m) \times \C H_{el}$}. \\
The tensor product $|g_1>\otimes |g_2>$ of two defects is to be
interpreted as ``a configuration with defect $g_1$ to the left of
defect $g_2$''. The order is important: if we measure the total flux
of $|g_1>|g_2>$ we get $g_1g_2$, while the total flux of
$|g_2>|g_1>$ is $g_2g_1$. Thus we define the action of the
projection operators on the tensor product as follows:
\begin{equation}
P_h (|g_1>\otimes |g_2>) = \delta_{h,g_1g_2}(|g_1>\otimes |g_2>).
\end{equation}

We now give a couple of examples of nematic phases and their
associated modified quantum double:

\begin{itemize}

\item{A chiral tetrahedral nematic} \\
A tetrahedral nematic has internal symmetry $H_{el}=T$, where $T$
is the tetrahedral group. There are no reflections in $H$ because
the phase is chiral. The magnetic group is
$H_m=\Pi_1(SO(3)/T)=\tilde{T}$, where $\tilde{T}$ is the double
cover of $T$ in $SU(2)$. Therefore the relevant modified quantum
double is $\Acal=F(\tilde{T})\times T$. If we are considering a
quantum mechanical nematic, and we have spinors  around, then
$\Acal=F(\tilde{T})\times \tilde{T}\equiv D(\tilde{T})$. This is
the quantum double of $\tilde{T}$. \\
\item{An achiral tetrahedral nematic} \\
Now $H_{el} = T_d$, the group of symmetries of a tetrahedron
including reflections. The magnetic group is the same as in the
chiral case: $H_m = \Pi_1(O(3)/T_d)=\Pi_1(SO(3)/T)=\tilde{T}$.
Therefore $\Acal=F(\tilde{T})\times T_d$. If we allow for spinor
electric irreps, then\footnote{We actually need to define
$\tilde{T_d}$, because $T_d$ is a subgroup of $O(3)$, not of
$SO(3)$. We get double covers of subgroups of $SO(3)$ using the
canonical two-to-one homomorphism from $SU(2)$ to $SO(3)$.
Exploiting the fact that $O(3) \simeq SO(3)\times\Z_2$, we can
lift $T_d$ to a subgroup of $ SU(2)\times\Z_2$.}
$\Acal=F(\tilde{T})\times \tilde{T_d}$.\\
Note that the algebra multiplication is determined by the action of
the electric group on the magnetic group, which we still need to
calculate. We do this in the next section.
\item{A uniaxial nematic} \\
The local symmetry is $H_{el}=SO(2)\rtimes\Z_2$. The magnetic groups
is $\Pi_1(SO(3)/SO(2)\rtimes\Z_2)\simeq\Z$, but the analysis is
actually more subtle: the relevant modified quantum double turns out
to be
\begin{equation}
\Acal=F(\widetilde{SO(2)\rtimes\Z_2})\times\C SO(2)\rtimes\Z_2.
\end{equation}
where $\widetilde{SO(2)\rtimes\Z_2}$ is the cover of $SO(2) \rtimes
\Z_2$ in $SU(2)$. The defects carry a continuous label. We stated
before that $H_m$ is always discrete, and indeed in this case
$\Pi_1(SO(3)/SO(2)\rtimes\Z_2)\simeq \Z_2$. Thus there is only one
nontrivial defect homotopy class. Given a configuration of the
fields that corresponds to this defect, we can always act on the
configuration with the residual rotational symmetries of
$SO(2)\subset H_{el}$ to obtain a continuous family of
configurations that correspond to the same defect. Because the
defects in this family all have the same energy, we should expect a
zero mode in this defect sector. This in turn leads to interesting
phenomena, such as Cheshire charge (for a discussion, in a two
dimensional context, see references \cite{Bais:2003mx}). Cheshire
charge is usually associated with gauge theories, but it can exist
in global theories as well\cite{McGraw}. We point out that the Hopf
symmetry approach is applicable to these cases, though beyond the
scope of this paper.
\end{itemize}

\subsubsection{Dyonic modes}

We now want to complete our description of the full ``internal''
Hilbert space by including the mixed sectors carrying both
nontrivial electric and magnetic charges. For every defect class $A$
we choose a \emph{preferred element} $g_A$ as a representative, then
all defects can be written as $h\cdot g_A$ for some $A$ and some
$h\in H_m$. Call $N_A$ the normalizer\footnote{The normalizer of
$g_A$
 is the subgroup of $H_{el}$ whose elements $h$ satisfy
 $h\cdot g_A = g_A$.} of $g_A$.
The normalizers of elements in the same defect class are
isomorphic:
$N_{h\cdot g_A}=h N_A h^{-1}$.\\
A dyonic mode is an electric mode in a topologically nontrivial
sector corresponding to a defect $g_A$. In that case there is an
important restriction due to the  topological obstruction to
globally implement all global symmetry transformations of $H_{el}$
\cite{Bala,dwpb1995}. Only  the subgroup $N_A$ can be globally
implemented, and hence the electric modes in such a sector will
transform under an irrep $\alpha$ of
$N_A$. \\
Extending our ket notation to all dyonic/mixed  sectors we denote a
state with electric component $|e^{\alpha}_j>$ and a defect $h\in A$
in the background by $|h,e^{\alpha}_j>$ (following the notation in
the literature \cite{Dijkgraaf}). The $|e^{\alpha}_j>$ form a basis
of the vector space on which $\alpha$ acts, so that the
$|h,e^{\alpha}_j>$ (with $h \in A$) are a basis of the vector space
associated to the dyon. We denote this irrep of our dyon by
$\Pi^A_{\alpha}$.\\
The action of global symmetry transformations on this vector space
is subtle. If we take a transformation $n\in N_A$, then
\begin{equation}
n\cdot|h,e^{\alpha}_j> \equiv \Pi^A_{\alpha}(n)|h,e^{\alpha}_j> =
|h,\alpha(n)e^{\alpha}_j>,
\end{equation}
i.e. $n$ acts on the electric mode.\\
But if the transformation $g\notin N_A$, it will transform the
defect, while at the same time it can act on the electric mode! To
describe this action, it is convenient to define another notation
for the
vectors in $\Pi^A_{\alpha}$. \\
First note that the elements of the defect class $A$ are in
one-to-one correspondence with left $N_A$-cosets in $H_{el}$.
Choose representatives $x^A_i$ of left $N_A$ cosets, such that
$x^A_1 = e$. Then $x^A_i$ corresponds to $h^A_i= x^A_i \cdot g_A$,
where $h^A_i$ is an element of $A$. This association is well
defined because it is independent of the particular choice of
representative $x^A_i$ of the left $N_A$ coset, since by
definition the elements of $N_A$ commute with $g_A$. Furthermore,
different $x^A_i$ correspond to different elements $h^A_i$ of $A$,
and we have \mbox{ $A = \{h^A_1=g_A, h^A_2, h^A_3,\ldots,h^A_n\}$
}. Now a basis of the vector space on which $\Pi^A_{\alpha}$ acts
is given by
$\{|h^A_i,e^{\alpha}_j>\}$. \\
Alternately, we can denote $|h^A_i,e^{\alpha}_j>$ by $x^A_i\tens
|e^{\alpha}_j>$. In this notation, acting on the defect $h^A_i$
with $g$ corresponds to \emph{multiplying} $x^A_i$ by $g$ from the
left. Thus
\begin{equation*}
g\cdot|h^A_i,e^{\alpha}_j> = gx^A_i\tens  |e^{\alpha}_j> = x^A_k
n\tens  |e^{\alpha}_j> \equiv |h^A_k,\alpha(n)e^{\alpha}_j>
\end{equation*}
where $gx^A_i = x^A_k n$, with $n\in N_A$. In other words,
$gx^A_i$ sits in some left $N_A$ coset. Since the $x^A_k$ form
representatives of left $N_A$ cosets, $gx^A_i$ is equal to $x^A_k
n$ for some $k$ and some $n\in N_A$. This $n$ then acts on the
electric part of the dyon. This notation is the most transparent
notation we can adopt for the action of $g$ on the dyon. \\
The action of the projection operator $P_h$ on the dyon is
\begin{equation}
P_h\cdot|h^A_i,e^{\alpha}_j> =
\delta_{h,h^A_i}|h^A_i,e^{\alpha}_j>,
\end{equation}
thus it projects the defect part.\\
Summarizing, the $\Pi^A_{\alpha}$ are irreps of $F(H_m) \times \C
H_{el}$. It turns out that these are \emph{all} the
irreps\footnote{This follows because $F(H_m) \times \C H_{el}$ is
a transformation group algebra. We can then use a theorem
described elsewhere\cite{Bais:2002ny}.} of $F(H_m) \times \C
H_{el}$. We denote the vectors on which $\Pi^A_{\alpha}$ acts by
$V^A_{\alpha}$.

Note that the electric and magnetic modes discussed are also irreps
of $F(H_m) \times \C H_{el}$. Namely, electric modes are irreps
$\Pi^{C_e}_{\alpha}$, with $C_e$ the conjugacy class of the identity
$e$: $C_e=\{e\}$, and the excitations carry irreps of the full
group, i.e. $N_A=H_{el}$.  Magnetic modes are irreps $\Pi^A_1$
(where $1$ is the identity or trivial representation), dyons with a
trivial representation of $N_A$. In this sense \emph{the quantum
double offers a unified
description of electric, magnetic and dyonic modes}. \\

The steps towards classifying all the irreps of the double
$F(H_m)\times\C H_{el}$ are relatively straightforward
\cite{Bais:2002ny}:
\begin{itemize}
\item Determine the defect classes $A$ of $H_m$, i.e. the classes
under the action of $H_{el}$
\item Pick a preferred element $g_A$ for every $A$, and determine
the normalizer $N_A$ of $g_A$
\item Determine the irreps $\alpha$ of $N_A$
\item The irreps of $F(H_m) \times \C H_{el}$ are the set
$\{\Pi^A_{\alpha}\}$.
\end{itemize}

\subsection{The Hopf symmetry description of achiral non-Abelian nematics}

\subsubsection{General aspects}

We haven't shown that all properties of electric, magnetic and
dyonic excitations are captured by $F(H_m)\times \C H_{el}$. For
example, we would like $F(H_m) \times \C H_{el}$ to reproduce the
fusion rules of these modes. This can be done, by introducing the
\emph{coproduct} $\Delta$ which in turn determines the tensor
products of the irreps. This works as follows: $\Delta$ is a map
from $F(H_m)\times \C H_{el}$ to $F(H_m)\times \C H_{el}\otimes
F(H_m)\times \C H_{el}$, that respects the multiplication (i.e.
it's an algebra morphism):
\begin{equation}
\Delta(P_h g P_{h'}g')=\Delta(P_h g)\Delta(P_{h'}g')
\end{equation}
Given an element $P_hg$, the coproduct can be written out in a
basis of $F(H_m)\times \C H_{el}\otimes  F(H_m)\times \C H_{el}$:
\begin{equation}
\Delta(P_h g) = \sum_{h_1,h_2\in H_m, g_1,g_2\in
H_{el}}\lambda_{h_1,g_1,h_2,g_2} P_{h_1}g_1\otimes  P_{h_2}g_2.
\nonumber
\end{equation}
Because this is rather cumbersome notation, we adopt the more
convenient \emph{Sweedler's notation} instead. For any $a\in
F(H_m)\times \C H_{el}$ we write
\begin{equation}
\Delta(a)=\sum_{(a)} a^{(1)}\otimes  a^{(2)}.
\end{equation}
This means we can write $\Delta(a)$ as a sum of elements of the
form $a^{(1)}\otimes  a^{(2)}$,
with $a^{(1)}$ and $a^{(2)} \in F(H_m)\times \C H_{el}$.\\
Now if we have two irreps $\Pi^A_{\alpha}$ and $\Pi^B_{\beta}$,
their tensor product is a representation of $F(H_m)\times \C
H_{el}$ whose action on $a$ is given by
\begin{equation*}
(\Pi^A_{\alpha} \otimes  \Pi^B_{\beta}) a \equiv (\Pi^A_{\alpha}
\otimes  \Pi^B_{\beta})\Delta(a) =
\sum_{(a)}\Pi^A_{\alpha}(a^{(1)}) \otimes  \Pi^B_{\beta}(a^{(2)}).
\end{equation*}
It is possible to choose the coproduct in such a way that it
produces the right fusion rules. The answer is
\cite{dwpb1995,bmbreaking:2006}
\begin{equation}
\Delta(P_h g) = \sum_{h'\in H_m} P_{hh'^{-1}} g \otimes  P_{h'}g.
\end{equation}
$F(H_m)\times \C H_{el}$ is also a \emph{Hopf algebra}. This means
there is even more structure on $F(H_m)\times \C H_{el}$ than we
have defined until now.
Here we will only introduce the structures that are relevant for this chapter and the next.\\
A Hopf algebra has a \emph{counit} $\eps$, which corresponds to the
\emph{trivial or vacuum representation}. For the case of
$F(H_m)\times \C H_{el}$, $\eps$ is defined by
\begin{equation}
\eps(P_hg) = \delta_{h,e}.
\end{equation}
One may also consider $\eps$ as a one-dimensional representation
of the double $F(H_m)\times \C H_{el}$, whose tensor product with
any irrep $\Pi^A_{\alpha}$ gives $\Pi^A_{\alpha}$:
\begin{equation}
\Pi^A_{\alpha}\otimes  \eps \simeq \eps\otimes \Pi^A_{\alpha}
\simeq \Pi^A_{\alpha}.
 \end{equation}
We introduce one more structure: the \emph{antipode} $S$, defined
for $F(H_m)\times \C H_{el}$ by
\begin{equation}
S(P_hg)=P_{g^{-1}\cdot h^{-1}}g^{-1}.
\end{equation}
It is used to define the \emph{conjugate or antiparticle
representation} $\overline{\Pi^A_{\alpha}}$ of $\Pi^A_{\alpha}$:
\begin{equation}
\overline{\Pi^A_{\alpha}}(P_hg) = ({\Pi^A_{\alpha}} S(P_hg))^t,
\end{equation}
where t denotes the transpose. The properties of the antipode
imply that this $\overline{\Pi^A_{\alpha}}$ is a representation,
and that the vacuum representation $\eps$ appears in the
decomposition of $\overline{\Pi^A_{\alpha}}\otimes
\Pi^A_{\alpha}$:
\begin{equation}
\overline{\Pi^A_{\alpha}}\otimes \Pi^A_{\alpha}= \eps\oplus
\bigoplus_{B,\beta}\Pi^B_{\beta}.
\end{equation}
This property explains the term ``antiparticle irrep'': an irrep and
its anti-irrep can ``annihilate'' into the vacuum representation
$\eps$ (i.e. there is no topological obstruction to such a decay).
This discussion applies to general Hopf algebras, and therefore to
any physical system characterized by a Hopf symmetry.

\subsubsection{Braiding and quasitriangular Hopf algebras}

Braiding plays a crucial role in the double symmetry breaking
formalism. We have discussed braiding as it features in the
present context in some detail elsewhere \cite{bmmelting:2006},
therefore we will be brief here. We are especially interested in
the way braiding is implemented in the algebraic structure of a
modified quantum double. First we review the case where the Hopf
symmetry $\Acal$ is a quantum double $D(H)$, which is a modified
quantum double with $H_{el} = H_m = H$. Then we address the case
where $\Acal = F(H_m)\times \C H_{el}$, with $H_m \neq H_{el}$.

Braiding addresses the following question: What happens to a
two-particle state when one excitation is adiabatically (i.e.
slowly) transported around the other? The braiding properties are
encoded in the \emph{braid operator} $\braid$. When two defects
$|g>$ and $|h>$ ($g,h \in H_m=H$) are braided, it is known what the
outcome is (it follows from homotopy
theory\cite{Poenaru,Mermin,Bais:1980fm}). If $|g>$ lives in $V^A$
and $|h>$ in $V^B$, then $\braid$ is a map from $V^A\otimes  V^B$ to
$V^B\otimes  V^A$ whose action is defined by
\begin{equation}
\braid\cdot |g> \otimes|h> = |ghg^{-1}> \otimes|g>
\label{eq:braiddefs}
\end{equation}
Note that it braids the defect to the right halfway around the
other defect, and we call this a \emph{half-braiding}. To achieve
a full braiding, or monodromy we have to apply $\braid^2$.\\
The equation for the braiding of defects $|g>$ and $|h>$ we have
just discussed applies equally well to the case of global as to the
case of gauged symmetry.

Electric modes braid trivially with each other\footnote{Actually, if
we are braiding two indistinguishable electric particles, then the
wave function of the system picks up a phase factor $e^{i2\pi s}$,
where $s$ is the spin of the particles. In the phases discussed in
this article, the electric modes are spinless.}:
\begin{equation}
\braid |v_1> \otimes |v_2> = |v_2> \otimes |v_1>.
\label{eq:braidelel}
\end{equation}
The (full) braiding of an electric mode with a topological defect
leads to the phase factor causing the famous Aharonov-Bohm effect.
In the present non-Abelian context that means that if we carry a
particle in a state $|v>$ of a representation $\alpha$ of the group
$H$ adiabatically around a defect with topological charge $g \in H $
then that corresponds to acting with the element $g$ in the
representation $\alpha$ on $|v>$:
\begin{equation}
\braid |g>\otimes |v> = \alpha(g)|v> \otimes|g>.
\label{eq:braidelmag}
\end{equation}
In the global case, during the parallel transport the particle is
following a curved path in its internal space. It is being ``frame
dragged'', as it is called \cite{Wilczek}. To be specific, one
defines a local coordinate frame somewhere at the start of the path
in $G$ characterizing the defect. Then one lets the elements of the
path act on this initial frame, to obtain a new frame everywhere on
the path. An electric mode is then parallel transported around the
defect if its coordinates are constant with respect to the local
frames. This is basically the reason that one obtains the same
outcome for parallel transport as in a local gauge theory, the only
difference being that in a gauged theory the particular path in
Hilbert space taken by the electric mode is gauge dependent and
therefore not a physical observable. Only its topological winding
number leads to an observable effect.

There is a continuum formulation of lattice defects in terms of
curvature and torsion sources in Riemann-Cartan geometry
\cite{Katanaev,Kleinert2}. In this geometrical approach one can also
explicitly evaluate the outcome of parallel transport of an electric
mode around a defect. This idea has been applied to quite a few
phases, such as superfluid helium, where the symmetry is also global
\cite{Khazan}. It has also been applied to uniaxial nematic liquid
crystals in the one constant approximation \cite{McGraw} (in the
absence of diffusion).

One of the advantages of introducing the Hopf symmetry $\Acal$
(which we take to be a quantum double for now) is that a Hopf
algebra is naturally endowed with a so called \emph{universal R
matrix} $R$. $R$ is an element of $\Acal\otimes \Acal$. It encodes
the braiding of states in irreps of $\Acal$: to braid two states,
$|\phi_1>$ in $\Pi_1$ and $|\phi_2>$ in $\Pi_2$, act with $R$ on
$|\phi_1>\otimes |\phi_2>$, and then apply the flip operator $\tau$
This composition is called the \emph{braid operator} $\braid$:
\begin{equation*}
\braid (|\phi_1>\otimes  |\phi_2>) = \tau\circ(\Pi_1\otimes \Pi_2)
\circ R \circ |\phi_1> \otimes |\phi_2>.
\end{equation*}
where the action of $\tau$ is just to flip any two vectors
$|\phi_1>$ and $|\phi_2>$:
\begin{equation}
\tau (|\phi_1>\otimes |\phi_2>) = |\phi_2>\otimes |\phi_1>.
\end{equation}
If $|\phi_1>$ is in the vector space $V_1$, and $|\phi_2>$ in $V_2$,
then $|\phi_1> \otimes  |\phi_2>$ is a vector in $V_1\otimes V_2$.
Then $\braid (|\phi_1>\otimes |\phi_2>)$ is a vector in $V_2\otimes
V_1.$

The universal R matrix is an invertible element of $\Acal\otimes
\Acal$, i.e. there is an $R^{-1}\in \Acal\otimes \Acal$ which
satisfies
\begin{equation}
RR^{-1}=R^{-1}R = 1\otimes  1.
\end{equation}
$\braid$ corresponds to braiding the particle on the right in a
counterclockwise fashion halfway around the particle on the left.
Using $R^{-1}$, we can define the inverse braiding, which is the
clockwise braiding of the particle on the right halfway around the
particle on the left:
\begin{equation}
\braid^{-1} = R^{-1}\circ\tau.
\end{equation}
It is sometimes convenient to write $R$ in Sweedler's notation:
\begin{equation}
R = \sum_{(R)}R^{(1)}\otimes  R^{(2)}.
\end{equation}
We can let $R$ act on n-particle states. To do this, We define
\begin{equation}
R_{ij}=\sum_{(R)}1\otimes \cdots\otimes  R^{(1)}\otimes
\cdots\otimes R^{(2)}\otimes \cdots\otimes  1
\end{equation}
where $R^{(1)}$ is in the i-th, and $R^{(2)}$ in the j-th position.
$R_{ij}$ implements the half-braiding of particles $i$ and $j$. $i$
needn't be smaller than $j$. For example, on a two particle state
$R_{21}=\sum_{(R)}R^{(2)}\otimes  R^{(1)}$.

For the $D(H)=F(H)\times\C H$ case, the universal R matrix is given
by
\begin{equation}
R=\sum_{g\in G}P_g e \otimes  g.
\end{equation}
The braid operator $\braid$ that is derived from this $R$
reproduces the braiding of the different modes discussed in this
section.

The universal R matrix satisfies the \emph{Yang-Baxter equation}:
\begin{equation}
R_{12}R_{13}R_{23}=R_{23}R_{13}R_{12}.
\end{equation}

\begin{figure}[htbp]
  \begin{center}
  \includegraphics[height=3cm]{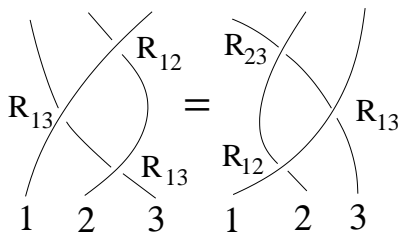}
    \caption{{\small The Yang-Baxter equation.}}
  \end{center}
\label{fig:YBE}
\end{figure}

A Hopf algebra with a universal R matrix is called a
\emph{quasitriangular Hopf algebra}, and the quantum double $D(H)$
is quasitriangular.

So far we have discussed braiding for the quantum double of a
discrete group, but we are also interested in the case where $H_m
\neq H_{el}$, and then we need to know what the outcome is of
braiding a vector $|v>$ around a defect $|g>$ (the braiding of
defects is the same as above, see (\ref{eq:braiddefs})). In the case
of non-Abelian nematics (and many other cases), the vector $|v>$ is
acted on by some element of $H_{el}$, and this element is
independent of the vector $|v>$. In other words, there is a map
$\Gamma: H_m \mapsto H_{el}$ that sends $g\in H_m$ to $\Gamma(g)$,
which acts on $|v>$ when $|v>$ is parallel transported around $|g>$.
$\Gamma$ is a group homomorphism, and is dictated by the physics of
the system we are considering. In the cases we have studied (in
particular the cases relevant for this article), $\Gamma$ also
satisfies the following relations:
\begin{eqnarray}
&&\forall g, g_1, g_2 \in H_m, h \in H_{el}:\\
&& \Gamma(g_1)\cdot g_2 = g_1 g_2 g_1^{-1} \label{eq:Gammaone}\\
&& \Gamma(h\cdot g) = h\Gamma(g)h^{-1}. \label{eq:Gammatwo}
\end{eqnarray}

$F(H_m)\tens \C H_{el}$ is then a quasitriangular Hopf algebra
with the following braid matrix:
\begin{equation}
R=\sum_{g\in H_m}P_g e \otimes  \Gamma(g).
\end{equation}
This equation precisely encodes what we have described above. The
inverse of $R$ is
\begin{equation}
R^{-1}=\sum_{g\in G}P_g e \otimes  \Gamma(g^{-1}).
\end{equation}

The quantum double $D(H)$ is a special case of a generalized quantum
double, with $H_m=H_{el}=H$, $h\cdot g = hgh^{-1}\;\forall h,g\in
H$, and $\Gamma\equiv id$, the identity operator. We will see
examples of phases with
nontrivial $\Gamma$ later on.\\

\subsubsection{Achiral non-Abelian nematics}
We will now explicitly describe the Hopf symmetry relevant for
non-Abelian nematics with tetrahedral, octahedral and icosahedral
residual symmetry. The Hopf symmetry $\Acal$ is of the form
discussed above: $\Acal = F(H_m)\times \C H_{el}$. We will
explicitly describe the electric and magnetic groups, and the
defect classes. We also analyze the consequence of the presence of
inversions and
reflections in $H_{el}$.\\

\textit{{\small Achiral tetrahedral nematic}} \\

\begin{figure}[t]
\centering
\includegraphics[scale=0.2,keepaspectratio=true]{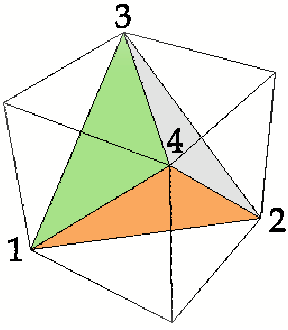}
\caption{{\small The tetrahedron, with labelled vertices.}}
\label{fig:tetverts}
\end{figure}

The electric group is $H_{el} = T_d$, the group of symmetries of a
tetrahedron, including reflections (since the phase is achiral),
see figure \ref{fig:tetverts}. We denote elements of $T_d$ as
permutations of the four vertices of a tetrahedron, e.g. (12),
(134), (13)(24), etc.

Before we can define the magnetic group, we first describe a
common parameterization of $SU(2)$, and the two-to-one
homomorphism from $SU(2)$ to $SO(3)$. To specify a rotation in
$SO(3)$, one specifies an axis around which the rotation takes
place, and a rotation angle $-\pi < \te \leq \pi)$. Denote by
$\hat{n}$ a unit vector along the axis of rotation, and define
positive $\te$ to correspond to counterclockwise rotation with
respect to $\hat{n}$. Then this rotation is denoted by
$R(\hat{n},\te)$. In this parameterization we can envisage $SO(3)$
as a ball of radius $\pi$ with antipodal points on the surface
identified.

We can parameterize matrices in $SU(2)$ in a very similar way: take
any unit vector $\hat{n}$, and any angle $-2\pi < \te \leq 2\pi)$.
Notice how $\te$ runs over a larger range than in the $SO(3)$ case,
we now have a sphere with radius $2\pi$ and the surface of the
sphere corresponds to the center element $-1$). Now associate to
$\hat{n}$ and $\te$ the following matrix in $SU(2)$:
\begin{equation}
 u(\hat{n},\te)= exp( i \frac{\te}{2} \hat{n}\cdot \vec{\sigma})=
 cos(\frac{\te}{2}) \unitmatrix + i
sin(\frac{\te}{2}) \hat{n}\cdot \vec{\sigma}
\end{equation}
where $\vec{\sigma} = ( \sigma_x , \sigma_y, \sigma_z)$ is a vector
containing the three Pauli matrices, and $\unitmatrix$ is the unit
matrix.
\begin{equation}
    \sigma_x =  \left( \begin{array}{cc}
                0 & 1 \\
                1 & 0
                \end{array} \right),
    \sigma_y =  \left( \begin{array}{cc}
                0 & -i \\
                i & 0
                \end{array} \right),
    \sigma_z =  \left( \begin{array}{cc}
                1 & 0 \\
                0 & -1
                \end{array} \right)
\end{equation}
The homomorphism $\rho: SU(2) \mapsto SO(3)$ is now easily
accomplished: replace $u$ by $R$.

In the achiral tetrahedral phase, the magnetic group is $\Td$:
\begin{equation*}
\Pi_1 (O(3)/T_d)=\Pi_1 (SO(3)/T)=\Pi_1 (SU(2)/\Td)\simeq \Td.
\end{equation*}
$\Td$ is the inverse under $\rho$ of $T$, the tetrahedral group in
$SO(3)$. Global symmetry transformations act on the defects in
$\Td$, so the defects are grouped together in orbits under the
action of the electric group, in this case $T_d$. We note that
this is not the same as the conjugacy classes of $\Td$! To see
this, we must first fix a notation for the defects. Write an
element of $\Td$ as
\begin{eqnarray}
u(\hat{n},\te) = sgn( cos \frac{\te}{2}) u'( \hat{n},\te' ) &&
- \pi < \te' \leq \pi \nonumber \\
&& \te' = \te (mod 2 \pi).
\end{eqnarray}
 Every defect corresponds to a $u(\hat{n},\te)$ (the endpoint of the
  path in $SU(2)$ that characterizes the defect).
We denote the defects as cycles of $T$ with square brackets with a
plus or minus sign, e.g. $\pm[123]$. The minus sign corresponds to
the $2\pi$ defect, i.e. the nontrivial loop in $SO(3)$. The defect
$[123]$ corresponds to the $u'(\hat{n},\te')$ with $- \pi < \te'
\leq \pi$ such that
\begin{equation}
R(\hat{n},\te') = (123).\nonumber
\end{equation}
We need to know the axis $\hat{n_1}$ in figure \ref{fig:tetverts},
such that $(123)$ corresponds to a $\frac{2\pi}{3}$ rotation
around $\hat{n_1}$.

From the figure we see that $\hat{n_1}=\frac{1}{\sqrt{3}}(1,1,1).$
Thus we define $[123] = u(\hat{n_1},\frac{2\pi}{3})$. The trivial
defect is denoted by $1$, and the $2\pi$ defect by $-1$.
\begin{table}[t]
\centering
\begin{tabular}{|c|l|}
\hline
 & \\
Preferred element $g_A$ &   Conjugacy class $A$\\
 & \\
\hline\hline
 & \\
$\pm e$                     &   $C_{\pm e}=\pm \{e\}$\\
 & \\
$\pm [123]$                 &   $C_{\pm [123]}=$ \\
& $\pm \{[123],[134],[142],[243]\}$\\
 & \\
$\pm [124]$                 &   $C_{\pm [124]}=$ \\
& $\pm \{[124],[132],[234],[143]\}$ \\
 & \\
$[(12)(34)]$            &   $C_{[(12)(34)]}=$ \\
& $\{\pm [(12)(34)],\pm [(13)(24)],$ \\
& $\pm [(14)(23)]\}$\\
 & \\
\hline
\end{tabular}
\caption{{\small The conjugacy classes of $\Td$, and the preferred
elements $g_A$. When we write $\pm$ in front of a class, we mean
that there are two classes, one with plus signs in front of all the
elements, and one with minus signs in front of all the elements. On
the other hand, if we write $\pm$ in front of an element inside a
class, we mean that the both the element with a plus and minus sign
in front of it are present in the class.}} \label{tab:tetconj}
\end{table}

To have a notation for all the defects in $\Td$, we first define
the following axes:
\begin{eqnarray}
\hat{n_1}=\frac{1}{\sqrt{3}}(1,1,1)&\quad&
\hat{n_2}=\frac{1}{\sqrt{3}}(1,1,-1)\nonumber \\
\hat{n_3}=\frac{1}{\sqrt{3}}(-1,1,1)&\quad&
\hat{n_4}=\frac{1}{\sqrt{3}}(-1,1,-1).\nonumber
\end{eqnarray}
Then the defects are given by $\pm e$ and
\begin{eqnarray}
\pm[123] =\pm u(\hat{n_1},\frac{2\pi}{3}) &&\pm[124] = \pm
u(\hat{n_2},\frac{2\pi}{3}) \nonumber\\
\pm[124]= \pm u(\hat{n_2},\frac{2\pi}{3}) &&\pm[234] =\pm
u(\hat{n_3},\frac{2\pi}{3})  \nonumber\\
\pm[134] = \pm u(\hat{n_4},\frac{2\pi}{3})&&\pm[132]= \pm
u(\hat{n_1},-\frac{2\pi}{3})  \nonumber\\
\pm[142] = \pm u(\hat{n_2},-\frac{2\pi}{3}) &&\pm[243] = \pm
u(\hat{n_3},-\frac{2\pi}{3}) \nonumber\\
\pm[143] = \pm u(\hat{n_4},-\frac{2\pi}{3}) &&\pm[(12)(34)]=\pm
u(\hat{z},\pi)  \nonumber\\
\pm[(13)(24)]=\pm u(\hat{y},\pi)  &&\pm[(14)(23)]=\pm
u(\hat{x},\pi). \nonumber
\end{eqnarray}

Using our notation, we can determine the defect classes under the
action of $T_d$. First consider an element $R(\hat{n},\phi)$ of
$T$. Its action on a defect $u(\hat{n_i},\te)$ gives
\mbox{$u(\hat{n},\phi)u(\hat{n_i},\te)u(\hat{n},\phi)^{-1}$}.

\begin{table}[t]
\centering
\begin{tabular}{|c|}
\hline
 \\
 Classes of $\Td$ under the action of $T_d$\\[3mm] \hline\hline
\\
$C_e$               \\[2mm]
$C_{-e}$            \\[2mm]
$C_{[123]}\cup C_{[124]}$\\[2mm]
$C_{-[123]}\cup C_{-[124]}$ \\[2mm]
$C_{[(12)(34)]}$\\
 \\
\hline
\end{tabular}
\caption{{\small The defect classes in a theory with $T_d$ symmetry.
The classes are unions of conjugacy classes of $\Td$. The conjugacy
classes of $\Td$ are defined in table \ref{tab:tetconj}.}}
\label{tab:tetd}
\end{table}

Now consider transformations in $T_d$ that are not connected to the
identity, such as the element $(12)$. These are called ``large''
symmetry transformations. We can always write a large symmetry
transformation as $Inv\times R(\hat{n},\phi)$, where $Inv$ is the
inversion matrix:
\begin{equation}\label{inv}
Inv = -\unitmatrix = \left( \begin{array}{ccc}
                                -1  &   0   &   0   \\
                                0   &   -1  &   0   \\
                                0   &   0   &   -1
                \end{array} \right).
\end{equation}
$Inv$ acts trivially on all the defects, thus the action of $Inv
\times R(\hat{n},\phi)$ on a defect $u(\hat{n_i},\te)$ gives
\mbox{$u(\hat{n},\phi)u(\hat{n_i},\te)u(\hat{n},\phi)^{-1}$}.

We will now write the large symmetry transformations as $Inv
\times R(\hat{n},\phi)$. First we define the  following axes:
\begin{eqnarray}
\hat{m}_1=\frac{1}{\sqrt{2}}(1,-1,0)\quad
\hat{m}_2=\frac{1}{\sqrt{2}}(-1,0,1)&&\nonumber \\
\hat{m}_3=\frac{1}{\sqrt{2}}(0,1,1) \quad
\hat{m}_4=\frac{1}{\sqrt{2}}(0,-1,1) && \nonumber \\
\hat{m}_5=\frac{1}{\sqrt{2}}(1,0,1) \quad
\hat{m}_6=\frac{1}{\sqrt{2}}(1,1,0). &&\nonumber
\end{eqnarray}

The inversions are given by
\begin{eqnarray}
(12) = Inv \times R(\hat{m}_1,\pi) & (13) = Inv\times R(\hat{m}_2,\pi) &  \nonumber \\
(14) = Inv \times R(\hat{m}_3,\pi) & (23) = Inv\times R(\hat{m}_4,\pi) & \nonumber \\
(24) = Inv \times R(\hat{m}_5,\pi) & (34) = Inv \times
R(\hat{m}_6,\pi) . \nonumber
\end{eqnarray}
We can now derive the multiplication table of $\Td$ using the
multiplication in $SU(2)$. For example,
\begin{eqnarray}
&& [123][123] = -[132]\nonumber\\
&& [123][124] = [(14)(23)]\nonumber\\
&& (12)[(12)(34)](12) = - [(12)(34)] \nonumber\\
&&\textrm{etc.} \nonumber
\end{eqnarray}

The Hopf symmetry of the achiral tetrahedral nematic is the
modified quantum double $F(\Td)\times\C T_d$. We just defined the
action of $T_d$ on $\Td$, which sets the multiplicative structure
of $F(\Td)\times\C T_d$.

Now that we have elucidated the action of $H_{el}$ on $H_m$, we can
determine the defect classes. These defect classes are the union of
conjugacy classes of $\Td$. The conjugacy classes of $\Td$ are shown
in table \ref{tab:tetconj}, and the defect classes in table
\ref{tab:tetd}. The ``small'' symmetry transformations (that are
connected to the identity) simply conjugate the defects, while the
large symmetry transformations may transform defects in different
conjugacy classes into each other.

\begin{table}[t]
\centering
\begin{tabular}{|c|c|c|}
\hline
 & & \\
$g_{\Orb}$          &       Centralizer $N_{\Orb}$ & Irreps\\
 & & \\
\hline\hline
&&\\
$\pm e$         &       $T$                             &$\Pi^{e}_{\alpha}$\\
 & & \\
$\pm [123]$ & $\{e,(123),(132)\}\simeq \Z_3$            &$\Pi^{\pm[123]}_k$, $k\in \Z_3$ \\
 & & \\
$\pm [124]$ & $\{e,(124),(142)\}\simeq \Z_3$            &$\Pi^{\pm[124]}_k$, $k\in \Z_3$\\
 & & \\
$[(12)(34)]$ & $\{e,(12)(34), (13)(24), (14)(23)\}$     &$\Pi^{\pm[123]}_k, k\in \Z_4$\\
            &          $ \simeq \Z_4$                   &                   \\
            & & \\
\hline
\end{tabular}
\caption{{\small The irreps of $F(\Td)\tens \C T_d$. The $\alpha$
are irreps of $T_d$.}} \label{tab:tet}
\end{table}

The centralizers and irreps of $F(\Td)\times\C T_d$ are given in
table \ref{tab:tet}.

We must carefully interpret the result that defect classes can be
larger than conjugacy classes. The defects in the same class have
the same energy, since the symmetry transformations commute with
the Hamiltonian. They also have isomorphic centralizers. However,
\emph{the cores of defects only related by a large symmetry cannot
be interchanged}. This is due to the fact that the symmetry isn't
connected to the identity. Thus these defects are \emph{not}
topologically equivalent: they cannot be deformed into each other
with a finite amount of energy. So we might conclude that we
shouldn't act with large symmetries on the defects, and we should
work with conjugacy classes. However, we cannot simply neglect the
large global symmetries, since they act on the electric modes.

Finally, we need a map $\Gamma$ from $H_m$ to $H_{el}$ to define a
braiding (see above): $\Gamma$ is given by
\begin{equation}
\Gamma(\pm[123])=(123),
\Gamma(\pm[(12)(34)])=(12)(34),etc.\nonumber
\end{equation}
Thus $\Gamma$ turns square brackets into round brackets and
neglects the eventual minus sign.\\
\vspace{0.5in}
\begin{table}[t]
\centering
\begin{tabular}{|c|l|}
\hline
 & \\
Pref. el. $g_A$ &   Defect class $A$ of $\Od$\\
 & \\
\hline\hline
 & \\
$\pm e$                     &   $C_{\pm e}=\{\pm e\}$\\
 & \\
$\pm[123]$                  &   $C_{\pm[123]}=\pm \{[123],[124],[142],[132],[134],$\\
                        &   $[234],[142],[143],[243]\}$         \\
 & \\
$\pm[1234]$             &   $C_{\pm[1234]}= \pm \{[1234],[1243],[1324],$ \\
                        &   $[1342],[1423],[1432]\}$        \\
 & \\
$[(12)(34)]$            &   $C_{[(12)(34)]}= \{\pm [(12)(34)],\pm
[(13)(24)] , \pm[(14)(23)]\}$\\
 & \\
$[12]$                  &   $C_{[12]}=\{\pm [12],\pm [13],\pm [23],\pm [14],\pm [24],\pm [34]\}$\\
& \\
\hline
\end{tabular}
\caption{{\small The defect classes of $\Od$ under the action $O_i$,
and the preferred elements $g_A$. See \ref{tab:tetconj} for a
description of the notation.}} \label{tab:octconj}
\end{table}

\textit{{\small Achiral octahedral nematic}}\\

The electric group is $O_i$, which is the octahedral group $O$
(consisting of the rotational symmetries of a cube), plus the
inversions and reflections of a cube. $O$ is isomorphic to $S_4$
(all permutations of the diagonals of a cube). Thus we can write
elements of $O$ as cycles, e.g. $(1234), (123), (23), (13)(24)$,
etc. Defining $Inv$ as in equation (\ref{inv}), we have that $O_i
= \{O, Inv\times O\}$. Thus elements of $O_i$ are (1234), (123),
Inv, $Inv \times (134)$, etc.

In analogy with the achiral tetrahedral nematic discussed above,
we can denote elements of the magnetic group $\Od$ as cycles with
square brackets, with an eventual minus sign. Examples are
$\pm[123], \pm[(12)(34)], \pm[(12)]$, where the minus sign is the
$2\pi$ defect.

The Hopf symmetry is $F(\Od) \times C O_i$. The defect classes in
$\Od$ under the action of $O_i$ are given in table
\ref{tab:octconj}. The inversion $Inv$ acts trivially on the
defects, so for the octahedral nematic the defect classes are in
fact the conjugacy classes of $\Od$. In general, when a group
carries a sublabel $i$, it means that the group contains $Inv$.

\begin{table}[t]
\centering
\begin{tabular}{|c|l|}
\hline
 & \\
Pr. el. $g_A$&   Defect class $A$ of $\Id$\\
 & \\
\hline\hline
 & \\
$\pm e$                     &   $C_{\pm e}=\pm \{e\}$\\
 & \\
$\pm[123]$              &   $C_{\pm[123]}=\pm\{[123],[132],[124],[142],[125],$ \\
                        &   $[152],[134],[143],[135],[153],$    \\
                        &   $[145],[154],[234],[243],[235],$    \\
                        &   $[253],[245],[254],[345],[354]\}$   \\
 & \\
$\pm[12345]$            &   $C_{\pm[12345]}= \pm\{[12345],[12453],[12534],[13254],$\\
                        &   $[13425],[13542],[14235],[14352],$\\
                        &   $[14523],[15243],[15324],[15432]\}$\\
 & \\
$\pm[12354]$            &   $C_{\pm[12354]} = \pm \{[12354],[12435],[12543],[13245],$   \\
                        &   $[13452],[13524],[14253],[14325],$                      \\
                        &   $[14532],[15234],[15342],[15423]\}$             \\
 & \\
$[(12)(34)]$            &   $C_{[(12)(34)]}=$                               \\
                        & $\{\pm[(12)(34)], \pm[(12)(35)], \pm[(12)(45)], \pm[(13)(24)], $ \\
                        & $\pm[(13)(25)], \pm[(13)(45)], \pm[(23)(45)], \pm[(14)(23)], $\\
                        & $\pm[(14)(25)], \pm[(14)(35)], \pm[(24)(35)], \pm[(15)(23)], $\\
                        & $\pm[(15)(24)], \pm[(15)(34)], \pm[(25)(34)]\}$\\
 & \\
$\pm[123]$              &   $C_{\pm[123]} = \pm\{[123],[124],[125],[132],[134],$\\
                        &   $[234],[135],[235],[142],[143],$\\
                        &   $[243],[145],[245],[345],[152],$\\
                        &   $[153],[253],[154],[254],[354]\}$\\
 & \\
\hline
\end{tabular}
\caption{{\small The defect classes of $\Id$ under the action $I_i$,
and the preferred elements $g_A$. See table \ref{tab:tetconj} for a
description of the notation.}} \label{tab:icosconj}
\end{table}
\vspace{0.2in}
\textit{{\small Achiral icosahedral nematic}}\\

The electric group is $I_i$, which consists of the icosahedral
group $I$ (the rotational symmetries of an icosahedron), plus
inversions and reflections of the icosahedron. $I$ is isomorphic
to $A_5$ (the even permutations of the five cubes inscribed inside
an icosahedron). Thus we write elements of $I$ as cycles, e.g.
(12345), (14)(23), etc. Just as in the octahedral case, $I_i  =
\{I, Inv\times I\}$.

We can denote elements of the magnetic group $\Id$ as cycles with
square brackets, with an eventual minus sign. Examples are
$\pm[123], \pm[(12)(345)], \pm[(15)]$, where the minus sign is the
$2\pi$ defect.
The Hopf symmetry is $F(\Id)\times\C I_i$. The defect classes in
$\Id$ under the action of $I_i$ are given in table
\ref{tab:icosconj}. The inversion $Inv$ acts trivially on the
defects, so the defect classes are the conjugacy classes of $\Id$.

\section{Defect condensates and Hopf symmetry breaking}

\subsection{The Hopf symmetry breaking formalism}

Using the Hopf symmetry description of a phase, one can study
phase transitions induced by the condensation of any mode, be it
electric, magnetic or dyonic. The theory of Hopf symmetry breaking
was first proposed by Bais, Schroers and Slingerland
\cite{Bais:2002pb,Bais:2002ny}. who developed the framework that
determines the physics of the broken phase, and applied it to
discrete gauge theories. The aim of the present article is to
apply this framework to non-Abelian nematics. It turns out that we
were physically motivated to alter one step of the framework,
namely the definition of the residual symmetry algebra. In this
section, we briefly introduce the Hopf symmetry breaking
formalism, relying strongly on physical motivation. The
mathematics of our approach, which differ slightly from the
mathematics in papers just mentioned, are discussed in another
paper \cite{bmbreaking:2006}.

Classically, a condensate corresponds to a nonzero expectation of
some (dis)order parameter field. This nonzero expectation value has
certain symmetries, its symmetry group is called the residual
symmetry group $H$. The quantum interpretation is that the ground
state of the system corresponds to a non-vanishing homogeneous
density of particles in some particular state. The residual symmetry
operators are the operators that leave that particular state
invariant. Whether we are working classically or quantum
mechanically, the residual symmetry operators are determined the
same way: determine the operators that act trivially on the
condensate, which is a vector in some irrep of the original
symmetry.

If the original symmetry was a group $G$, it would be easy to define
what we mean by the symmetry operators of a vector $|\phi_0>$ in an
irrep $\Pi$ of $G$, namely operators $h\in G$ that satisfy
\begin{equation}
h\cdot|\phi_0>=|\phi_0>.
\end{equation}
If the original symmetry is a Hopf algebra $\Acal$, on the other
hand, the definition of ``residual symmetry operators'' is not so
trivial. This is discussed in a related paper\cite{bmbreaking:2006}:
There is a physically very attractive definition of a ``residual
symmetry operator'': \emph{an operator is a residual symmetry
operator if its action on a particle state is not affected  by the
fusion of that particle with the state of the particles in the
condensate}. This means that if this operator acts on any particle,
and that particle fuses with the condensate, then the action of the
operator on the particle before and after fusion with the condensate
is the same. We have to make a choice at this point: whether the
particle fuses with the condensate from the left or the right. This
is an important point as we argue elsewhere\cite{bmbreaking:2006}.
Here we choose fusion with the condensate from the right. This leads
to the definition of $\Tright$, the \emph{right residual symmetry
algebra}, which consists of all symmetry operators in $\Acal$ whose
actions on a state are not affected by fusion of the state with the
condensate on the right. $\Tright$ is the analog of the residual
symmetry group when a group symmetry is spontaneously broken. We
note that there are cases where $\Tright$ isn't a Hopf algebra.

Once $\Tright$ has been established, we must take a close look at
the particles in this broken theory, which we now consider to be the
irreps of $\Tright$. It turns out that some particle species don't
braid trivially with the condensate. The presence of such particles
in the system implies a half-line discontinuity in the condensate,
across which the internal state of the condensate jumps, which means
that these particles have to be connected to a domain wall. Hence,
such particles will be \emph{confined}. We can similarly  determine
which particles are \emph{unconfined}, i.e. braid trivially with the
condensate, and (in the cases we've worked out) these particles turn
out to be the irreps of a new Hopf algebra called the
\emph{unconfined symmetry algebra} which will be denoted as
$\Ucal$.

\begin{table}[t]
\centering
\begin{tabular}{|c|c|c|c|}
\hline
 & & & \\
 Single defect condensate in &       &               &           \\
$\Acal =F(\Td)\tens\C T_d $ & $K$   &$\Rcal$        &   $\Ucal$ \\
 & & & \\
\hline\hline
 & & & \\
 $|-e>$         & $\tilde{C_1}$  &$F(T)\tens \C T_d$    & $F(T)\tens \C T_d$\\
 & & & \\
 $|[123]>$      & $\tilde{C_3}$  &$F(T/C_3)\tens \C C_3$& $D(e)$        \\
  & & & \\
$|[(12)(34)]>$ & $\tilde{C_2}$  &$F(T/C_2)\tens \C D_2$& $D(C_2)$      \\
 & & & \\
 $|-[(12)(34)]>$& $\tilde{C_2}$  &$F(T/C_2)\tens \C D_2$& $D(C_2)$      \\
 & & & \\
\hline
\end{tabular}
\caption{{\small Single defect condensates in a tetrahedral
nematic. $\Acal$ is the original Hopf symmetry, $K$ is the
smallest subgroup of $H_m$ that contains all the defects $g_i$
that are in the condensate, $\Rcal$ is the residual symmetry
algebra, and $\Ucal$ is the unconfined Hopf symmetry.}}
\label{tab:Tdsd}
\end{table}

Thus, in contrast with the conventional symmetry breaking analysis,
we have to distinguish two steps in the symmetry breaking: first to
the residual symmetry algebra $\Tright$, and then to the unconfined
symmetry algebra $\Ucal$. One should not make the mistake of
believing that $\Ucal$ contains all the information about the broken
phase: one might believe that the confined particles should be
neglected because they cost an infinite amount of energy to create
in a system of infinite extent, since the half-line discontinuity
costs a finite amount of energy per unit length. However, the
half-line may end on another confined excitation, thus giving rise
to a wall of finite length. We call a configuration consisting of
confined excitations connected by walls, such that the overall
configuration is unconfined, a \emph{hadronic composite}, in analogy
with hadrons in Quantum Chromodynamics, where hadrons are unconfined
composites of confined quarks. $\Tright$ contains all the
information about the hadrons, although it is not trivial to extract
this information\cite{bmbreaking:2006} experimentally.

\subsection{Defect condensates and residual symmetry algebras}

One can derive general formulae for $\Tright$ and $\Ucal$ in the
case of defect condensates in a phase with $F(H_m) \times \C H_{el}$
symmetry\cite{bmbreaking:2006}. Here we will just summarize the
resulting formulae, and in the next section apply them to the
classification of all possible defect condensates in achiral
tetrahedral, octahedral and icosahedral nematics.

\begin{table}[t]
\centering
\begin{tabular}{|c|c|c|c|}
\hline
 & & & \\
Class sum def. condensate  &           &                   &               \\
of $\Acal =F(\Td)\tens\C T_d $ & $K$       &$\Rcal$            &   $\Ucal$     \\
 & & & \\
\hline\hline
 & & & \\
 $|C_{-e}>$& $\tilde{C_1}$   &$F(T)\tens \C T_d$& $F(T)\tens \C T_d$       \\
  & & & \\
$|C_{[123]}>,|C_{-[123]}>,$   & $\Td$     &$\C T$ & $D(e)$\\
 $\qquad |C_{[124]}>,|C_{-[124]}>$  & & & \\
 & & & \\
$|C_{[(12)(34)]}>$& $\tilde{D_2}$   &$F(\Z_3)\tens \C T_d$& $F(\Z_3)\tens\C (\Z_3)_d$  \\
 & & & \\
\hline
\end{tabular}
\caption{{\small Conjugacy class sum defect condensates in an
achiral tetrahedral nematic. $(\Z_3)_d$ is isomorphic to the
permutation group of 3 elements.}} \label{tab:Tdclsum}
\end{table}

Given a phase with Hopf symmetry $\Acal=F(H_m) \times \C H_{el}$,
the condensate $|\phi_0>$ is a vector in some irrep
$\Pi^A_{\alpha}$ of $\Acal$. We must demand that $|\phi_0>$ braid
trivially with itself, otherwise the condensate is ill-defined
(the condensate itself would be filled with half-line
singularities). Thus we demand that
\begin{equation}
\braid \cdot |\phi_0> |\phi_0> = |\phi_0> |\phi_0>. \label{eq:TSB}
\end{equation}

Now let us take $|\phi_0>$ to be a defect condensate, which means it
is a vector in a magnetic irrep $\Pi^A_1$ ($1$ is the trivial irrep
of the normalizer $N_A$). There are different types of defect
condensates which we wish to analyze. A basis of the vector space on
which this irrep acts is given by $\{|g^A_i>\}$, where the $g^A_i$
are the different defects in $A$. We consider the following types of
condensates:
\begin{itemize}
\item Single defect condensate
\begin{equation}
|\phi_0>=|g^A_i>
\end{equation}
\item Class sum defect condensate
\begin{equation}
|\phi_0>=\sum_{g_i \in A} |g_i> =: |C_{g_A}>
\end{equation}
where $C_{g_A}$ is a defect class (i.e. the orbit of $g_A$ under the
action of $H_{el}$). We denote the condensate by $|C_{g_A}>$, where
$g_A$ is the preferred element of $A$. \item Combined defect
condensate
\begin{equation}
|\phi_0>=\sum_{g_i\in E} |g_i>
\end{equation}
where $E$ is a subset of the defects in one class. We need only take
the elements to be within one class because it turns out that we
need only study the cases where the condensate is the sum of vectors
in the same irrep (all other phase transitions can be interpreted as
a sequential condensation of vectors in different
irreps\cite{bmbreaking:2006}.).
\end{itemize}

\begin{table}[t]
\centering
\begin{tabular}{|c|c|c|c|}
\hline
 & & & \\
Combined def. cond.  &     $K$ &   $\Rcal$     &       $\Ucal$ \\
of $\Acal =F(\Td)\tens\C T_d $     &       &               &           \\
 & & & \\
\hline\hline
 & & & \\
$|[123]>+|[132]>$& $\tilde{C_3}$ &$F(T/C_3)\tens \C C_{3v}$     &   $\C C_{1v}$ \\
 & & & \\
$|[(12)(34)]>+|[(13)(24)]>$&$\tilde{D_2}$&$F(T/D_2)\tens\C C_{2v}$&  $\C C_{1v}$   \\
 & & & \\
$\{|[(12)(34)]>+$&$\tilde{D_2}$&$F(T/D_2)\tens\C D_{2d}$& $\C C_{1v}$\\
$|[-(12)(34)]>+|[(13)(24)]>$ & & & \\$
+|-[(13)(24)]>\}$ & &                       &           \\
 & & & \\
$|[(12)(34)]>+|[-(12)(34)]>$&$\tilde{C_2}$&$F(T/C_2)\tens\C D_{2d}$& $F(C_2)$\\
 & & & $\tens\C C_{2v}$\\
 & & & \\
$|[(12)(34)]>+|[(13)(24)]>$&$\tilde{D_2}$&$F(\Z_3)\tens\C T$&  $D(\Z_3)$    \\
 $\qquad +|[(14)(23)]>$& & & \\
 & & & \\
\hline
\end{tabular}
\caption{{\small Combined defect condensates in an achiral
tetrahedral nematic (that satisfy trivial self braiding). No two
defect condensates have simultaneously the same $\Rcal$ and
$\Ucal$. Thus in principle the different defect condensates are
distinguishable.}} \label{tab:Tdcomp}
\end{table}

The single defect and class sum defect condensates are a special
case of combined defect condensate.

\subsubsection{Single defect condensate}

Condense $|g_A>$ in the magnetic irrep $\Pi^A_1$. The condensate
$|g_A>$ satisfies the trivial self braiding condition mentioned
above. The residual symmetry algebra is
\begin{equation}
\Rcal = F(H_m/(g_A)) \otimes\C N_A \label{eq:qdsingdef}
\end{equation}
where we define $(g_A)$ to be the smallest subgroup of $H_m$ that
contains $g_A$.

This result for $\Rcal$ has a very natural interpretation: the
residual electric group is $N_A$, the subgroup of $H$ that doesn't
conjugate the defect. The magnetic part $H/(g_A)$ is not necessarily
a group. It consists of left cosets of $(g_A)=\{\ldots
g_A^{-1},e,g_A,g_A^2,\ldots\}$. The defects are now defined modulo
the condensate defect $|g_A>$. In other words, if a particle in a
magnetic irrep of the residual symmetry $\Rcal$ fuses with the
condensate $|g_A>$, it is left unchanged. Thus its topological
charge is defined modulo $g_A$.

One can prove that $\Rcal$ is a Hopf algebra $\iff$ $(g_A)$ is a
normal subgroup of $H_m$ $\iff$ $H_m/(g_A)$ is a group.

The unconfined symmetry algebra is
\begin{equation}
\Ucal = F(\Gamma^{-1}(N_A)/(g_A))\times \C (N_A/\Gamma(g_A)).
\label{eq:Usinfdef}
\end{equation}
Note that we condensed $|g_A>$, where $g_A$ was a chosen defect in
the defect class $A$. $g_A$ was chosen arbitrarily, so our formulae
are general.

\begin{table}[t]
\centering
\begin{tabular}{|c|c|c|c|}
\hline
 & & & \\
Single def. cond.  &       &               &           \\
of $\Acal =F(\Od)\tens\C O_i $ & $K$   &$\Rcal$        &   $\Ucal$ \\
 & & & \\
\hline\hline
 & & & \\
 $|-e>$         & $\tilde{C_1}$  &$F(O)\tens \C O_i$            & $F(O)\tens \C O_i$\\
  & & & \\
$|\pm[12]>$        & $\tilde{C_2}$  &$F(O/C_2)\tens \C D_{2i}$     & $F(C_2)\tens \C C_{2i}$      \\
 & & & \\
  $|\pm[123]>$      & $\tilde{C_3}$  &$F(O/C_3)\tens \C C_{3i}$     & $\C C_i$      \\
 & & & \\
 $|\pm[1234]>$      & $\tilde{C_4}$  &$F(O/C_4)\tens \C C_{4i}$     & $\C C_i$      \\
  & & & \\
$|\pm[(12)(34)]>$  & $\tilde{C_2}$  &$F(O/C_2)\tens \C D_{4i}$     & $F(D_2)\tens \C D_{2i}$  \\
 & & & \\
\hline
\end{tabular}
\caption{{\small Single defect condensates in an achiral octahedral
nematic. When we write $\pm$ in front of a condensate, we mean that
the condensate with or without the minus sign gives the same
symmetry breaking analysis.}} \label{tab:octisd}
\end{table}

\subsubsection{Class sum defect condensates}

Condense the sum of the defects in the defect class $A$:
\begin{equation}
|\phi_0> = \sum_{g^A_i\in A} |g^A_i> =: |C_{g_A}>. \nonumber
\end{equation}
A class sum defect condensate satisfies the trivial self braiding
condition (\ref{eq:TSB}):
\begin{eqnarray}
\braid(|C_{g_A}>&\otimes& |C_{g_A}>) \nonumber \\
&=&\braid (\sum_{g^A_i\in A} |g^A_i>\otimes  \sum_{g^A_k\in A}
|g^A_k>)
\nonumber\\
&=& \sum_{g^A_i\in A}(\sum_{g^A_k\in A} |g^A_i g^A_k (g^A_i)^{-1}>) \otimes  |g^A_i> \nonumber\\
&=& \sum_{g^A_i\in A}(\sum_{g^A_k\in A} |g^A_k >) \tens  |g^A_i>.  \nonumber\\
&=& |C_{g_A}>\otimes |C_{g_A}> \nonumber
\end{eqnarray}
In going from the second to the third line, we use the fact that
$gAg^{-1}=A$ for any $g\in H_m$.

\begin{table}[t]
\centering
\begin{tabular}{|c|c|c|c|}
\hline
 & & & \\
Class sum def. cond.             &           &                   &               \\
of $\Acal =F(\Od)\tens\C O_i $ & $K$       &$\Rcal$            &   $\Ucal$     \\
 & & & \\
\hline\hline
  & & & \\
$|C_{[(12)(34)]}>$ & $\tilde{D_2}$  &   $F(D_3)\tens \C O_i $  & $F(D_3)\tens \C D_{3i}$      \\
 & & & \\
 $|C_{\pm[123]}>$       & $\Td$             &   $F(C_2)\tens \C O_i$   & $F(C_2)\tens \C C_{2i}$ \\
 & & & \\
 $|C_{\pm[1234]}>$      &   $\Od$           &   $\C O_i$                &   $\C C_i$    \\
 & & & \\
 $|C_{[12]}>$       &   $\Od$           &   $\C O_i$                &   $\C C_i$    \\
 & & & \\
\hline
\end{tabular}
\caption{{\small Conjugacy class sum defect condensates in an
achiral octahedral nematic. When we write $\pm$ is front of a
condensate, we mean that the condensate with or without the minus
sign gives the same symmetry breaking analysis.}}
\label{tab:octclsum}
\end{table}

A class sum condensate doesn't break the electric group at all!
Namely, all elements of $H_{el}$ act trivially on a defect class,
since for any $g\in H_{el}$ we have
\begin{eqnarray}
g\cdot |\phi_0> = g \cdot (\sum_{g^A_i\in A} |g^A_i>) &=&
\nonumber
\\\quad = \sum_{g^A_i\in A} |g\cdot g^A_i> &=& \sum_{g^A_i\in A} |g^A_i> =
|\phi_0>. \nonumber
\end{eqnarray}
Thus this condensate is invariant under all of $H_{el}$. For this
reason, such a condensate is admissible in a theory where the
symmetry is gauged and we call it a \emph{gauge invariant magnetic
condensate} (the condensate respects gauge invariance). In a global
theory all condensates are admissible.

The residual and unconfined symmetry algebras are respectively
\begin{eqnarray}
&&\Rcal= F(H_m/K) \tens  \C H_{el} \\
&&\Ucal= F(H_m/K)\tens \C H_{el}/\Gamma(K),
\end{eqnarray}
where $K$ is the smallest subgroup of $H_m$ that contains the
class $A$. From this definition, one can prove that $K$ is a
normal subgroup of $H_m$. Thus $H_m/K$ is a group, and $\Rcal$ is
a Hopf algebra.

Later on, we will consider \emph{conjugacy class sum defect
condensates}, i.e. condensates of a sum of defects in the same
conjugacy class of $H_m$. This can be a whole defect class, or it
can be smaller than a whole defect class. If it is smaller, then
the electric group is partially broken.

\subsubsection{Combined defect condensates}

Start with a phase with $F(H_m) \times \C H_{el}$ symmetry. Choose
an irrep $\Pi^A_{\alpha}$, and consider a condensate of the form
$\sum_{g_i \in E} |g_i>$, with $E$ a subset of the defects in one
defect class.

The demand of trivial self braiding (\ref{eq:TSB}) gives
\begin{eqnarray}
\braid(\sum_{g_i\in E}|g_i>\otimes \sum_{g_k\in E}|g_k>)&&=
\sum_{g_i\in E} |g_i> \otimes \sum_{g_k\in E}
|g_k>\nonumber\\
\iff&& \nonumber \\
\sum_{g_k\in E}\sum_{g_i\in E}|g_i g_k g_i^{-1}> \otimes |g_i>&&=
\sum_{g_i\in E} \sum_{g_k\in E}
|g_i> \otimes  |g_k> \nonumber \\
\iff&& \nonumber \\
\forall g_i \in E : \{ g_i g_k g_i^{-1}\}_{g_k \in E}&&=
\{g_k\}_{g_k\in E}. \label{eq:TSBcomp}
\end{eqnarray}

It is interesting in itself to study how many different defect
condensates satisfy this criterion. Defect-antidefect condensates
$|g>+|g^{-1}>$ always satisfy this criterion\footnote{Note that $g$
and $g^{-1}$ needn't be in the same conjugacy class.}, as do any set
of commuting elements in a certain conjugacy class, and class sum
defect condensates. The trivial self braiding condition plays a
crucial role in determining $\Rcal$\cite{bmbreaking:2006}.

The results for $\Rcal$ and $\Ucal$ are:
\begin{eqnarray}
&&\Rcal = F(H_m/K) \tens  \C M_E \nonumber \\
&&\Ucal = F(N_E/K)\tens \C M_E/\Gamma(K), \nonumber
\end{eqnarray}
where we must still define all the notation in these results.

\begin{table}[t]
\centering
\begin{tabular}{|c|c|c|c|}
\hline
 & & & \\
Combined def. cond.  &     $K$         &   $\Rcal$                     &       $\Ucal$ \\
of $\Acal =F(\Od)\tens\C O_i $     &               &                               &               \\
 & & & \\
\hline\hline
 & & & \\
$|[123]>+|[132]>$           &   $\tilde{C_3}$&   $F(O/C_3)\tens \C D_{3i}$  &   $\C C_{2i}$ \\
 & & & \\
$|[12]>+|[34]>$             &   $\tilde{D_2}$&   $F(O/D_2)\tens \C D_{4i}$  &   $F(C_2)\tens \C C_i$ \\
 & & & \\
$|[(12)(34)]> \qquad$ &   $\tilde{D_2}$&   $F(O/D_2)\tens \C D_{4i}$  &   $F(C_2)\tens \C C_{2i}$\\
 $\qquad +|[(13)(24)]>$& & & \\
 & & & \\
$|[12]>+|[13]>+|[23]>$      &   $\tilde{D_3}$&   $F(O/D_3)\tens \C D_{3i}$  &   $\C C_i$ \\
 & & & \\
\hline
\end{tabular}
\caption{{\small Combined defect condensates in an achiral
octahedral nematic.}} \label{tab:octcomb}
\end{table}

Define the following subset of $H_{el}$ (which needn't be a
subgroup):
\begin{equation}
V_E\subset H_{el}: \;V_E=\{x_i N_A\}_{g_i\in E} \label{eq:vedefs}
\end{equation}
where $N_A \subset H_{el}$ is the normalizer of the chosen preferred
element $g_A$ in $A$, and the $x_i$ satisfy $x_i g_A x_i^{-1}=g_i
\in E$. $V_E$ corresponds to the set of left $N_A$ cosets that
corresponds to the defects in the condensate (under a correspondence
discussed above).

Define the following subgroup of $H_{el}$:
\begin{eqnarray}
M_E\subset H_{el}:\; M_E &=& \{m \in H_{el}: \{m\cdot g_i\}_{g_i\in E} = \{g_i\}_{g_i\in E} \} \nonumber \\
                     &=& \{m\in H_{el}: m V_E = V_E\}.
\label{eq:MEdef}
\end{eqnarray}
$M_E$ is composed of the global symmetry transformations that
leave the condensate invariant.

\begin{table}[t]
\centering
\begin{tabular}{|c|c|c|c|}
\hline
 & & & \\
$\Acal =F(\Id)\tens\C I_i $ & $K$   &$\Rcal$        &   $\Ucal$ \\
 & & & \\
\hline\hline
 & & & \\
 $|-e>$         & $\tilde{C_1}$  &   $F(I)\tens \C I_i$         & $F(I)\tens \C I_i$\\
 & & & \\
 $|[123]>$      & $\tilde{C_3}$  &   $F(I/C_3)\tens \C C_{3i}$  & $\C C_i$  \\
 & & & \\
 $|-[123]>$     & $C_3$             &   $F(\Id/C_3)\tens \C C_{3i}$    & $\C C_i$  \\
 & & & \\
 $|\pm[(12)(34)]>$& $\tilde{C_2}$&   $F(I/C_2)\tens \C D_{2i}$  &   $F(C_2)\tens \C C_{2i}$ \\
  & & & \\
$|[12345]>$    &   $\tilde{C_5}$&   $F(I/C_5)\tens \C C_{5i}$  &   $\C C_i$    \\
 & & & \\
 $|-[12345]>$   &   $C_5$           &   $F(\Id/C_5)\tens \C C_{5i}$    &   $\C C_i$    \\
 & & & \\
\hline
\end{tabular}
\caption{{\small Single defect condensates in an achiral icosahedral
nematic.}} \label{tab:icosisd}
\end{table}

Also define
\begin{equation*}
N_E\subset H_m: \; N_E = \{ n \in H_m: \{ng_in^{-1}\}_{g_i\in
E}=\{g_i\}_{g_i\in E}\}.
\end{equation*}
Using (\ref{eq:Gammaone}): $\Gamma(g_1)\cdot g_2 = g_1 g_2
g_1^{-1}\;\forall g_1,g_2\in H_m$, we can prove that
\begin{equation}
\Gamma^{-1}(M_E) = N_E \quad \textrm{and}\quad \Gamma(N_E) = M_E.
\label{eq:MENE}
\end{equation}

Finally, we need one more definition:
\begin{equation}
K\subset H_m : \; K=(\{g_i\}_{g_i\in E}),
\end{equation}
where $(\{g_i\}_{g_i\in E})$ is the smallest subgroup of $H_m$ that
contains all the $g_i\in E$, i.e. the defects in the condensate.

The trivial self braiding equation (\ref{eq:TSBcomp}) implies that
$K\subset N_E$. Thus, according to (\ref{eq:MENE})  and
(\ref{eq:vedefs})
\begin{equation}
\Gamma(K)\subset M_E.\label{eq:Ktriv}
\end{equation}

Summarizing, the unconfined magnetic group is $N_E/K$, and the
unconfined electric irreps are those that have $\Gamma(K)$ in their
kernel, which means that the electric group is $ M_E/\Gamma(K)$. If
we take a quantum double $D(H)$ ($H=H_{el}=H_m$), the unconfined
symmetry algebra becomes $\Ucal = D(N_E/K)$, because in that case
$M_E = N_E$.

\subsection{Non-Abelian condensates in liquid crystals}

We are now in a position to apply the results obtained in the
previous sections to the case of non-Abelian nematics. We have
worked out pretty much exhaustive listings  of all possible phases
characterized by defect condensates in achiral tetrahedral,
octahedral and icosahedral nematics. In the corresponding tables
of defect condensates we give, $\Acal$ is the original Hopf
symmetry, $K$ is the smallest subgroup of $H_m$ that contains all
the defects $g_i$ that are in the condensate, $\Rcal$ is the
residual symmetry algebra, and $\Ucal$ is the unconfined Hopf
symmetry. The defect condensates satisfy trivial self braiding, as
we required above.

By looking at all the tables of defect condensates in this
section, we note that two different condensates in our tables
never give simultaneously the same $\Rcal$ and $\Ucal$. Some
condensates give the same unconfined symmetry algebra, but $\Rcal$
is then different\footnote{We have to be specific when we say that
different defect condensates give different answers. For example,
condensing $|[123]>$ or $|[124]>$ will give isomorphic answers,
which is why we don't both put them up in our table. However, the
$\Rcal$s are different isomorphic subalgebras of $\Acal$.}. Thus
there are differences in the spectrum, though these are often
hidden in the unconfined spectrum of the hadrons corresponding to
different condensates. These differences are therefore quite
subtle and may be hard to detect but in principle they are
distinguishable. The problem with measuring defect condensates,
for example, is that the conventional measuring techniques can
measure the electric symmetry group (by looking at Bragg
reflections, for example), but as far as we know there are no
techniques yet to measure the magnetic symmetry group. Naively
this amounts to identifying the surviving unconfined degrees of
freedom and their interactions, for example by certain
interference experiments. We would need to measure non-Abelian
statistics to probe the braiding properties of the particles in
the broken phase. Only recently have there been direct
measurements of fractional statistics \cite{Camino}. Nevertheless,
if suitable techniques were developed, then we could use our
tables to identify a plethora of new phases and determine which
condensates they correspond to.

\begin{table}[t]
\centering
\begin{tabular}{|c|c|c|c|}
\hline
 & & & \\
$\Acal =F(\Id)\tens\C I_i $ & $K$           &   $\Rcal$         &   $\Ucal$     \\
 & & & \\
\hline\hline
  & & & \\
$|C_{\pm[12345]}>$ &   $\Id$               &   $\C I_i$        &   $\C C_i$    \\
 & & & \\
 $|C_{\pm[12354]}>$ &   $\Id$               &   $\C I_i$        &   $\C C_i$    \\
  & & & \\
$|C_{[(12)(34)]}>$ &   $\Id$               &   $\C I_i$        &   $\C C_i$    \\
 & & & \\
\hline
\end{tabular}
\caption{{\small Conjugacy class sum defect condensates in an
achiral icosahedral nematic.} } \label{tab:icosclsum}
\end{table}

\subsubsection{Achiral tetrahedral nematic}

We have listed all defect condensates in an achiral tetrahedral
nematic: the single defect condensates are collected in table
\ref{tab:Tdsd}, the class sum condensates in table
\ref{tab:Tdclsum}, and the combined defect condensates in table
\ref{tab:Tdcomp} .

\subsubsection{Achiral octahedral nematic}

The single defect condensates breaking $F(\Od)\tens\C O_i$ are given
in table \ref{tab:octisd}. The class sum defect condensates are
given in table \ref{tab:octclsum}. Finally, the combined defect
condensates are given in table \ref{tab:octcomb}. We note that the
list presented here is very representative. The other conceivable
defect condensates give the same $\Rcal$ and $\Ucal$ as one of the
defect condensates shown here (except for a small difference: there
may be condensates where $K$ is actually the double of a $K$ given
here. That slightly changes the magnetic part of $\Rcal$, but
doesn't affect $\Ucal$). These other defect condensates are
trivially different from the ones in the table: for example, they
may be permutations of the of the numbers used in the naming of the
defects.
\begin{table}[t]
\centering
\begin{tabular}{|c|c|c|c|}
\hline
 & & & \\
$\Acal =F(\Id)\tens\C I_i $                 &    $K$     & $\Rcal$   & $\Ucal$      \\
 & & & \\
\hline\hline
 & & & \\
$|[(12)(34)]>+|[(13)(24)]>$                 &$\tilde{D_2}$   &   $F(I/D_2)\tens \C D_{2i}$  & $\C C_i$ \\
 & & & \\
$|[(12)(34)]+|[(13)(24)]>+$    &$\tilde{D_2}$   &   $F(I/D_2)\tens \C T_i$     & $\C C_{3i}$ \\
 $\qquad +|[(14)(23)]>$ & & & \\
 & & & \\
$|[(12)(34)]>+|[(12)(35)]>+$    &$\tilde{D_3}$   &   $F(I/D_3)\tens \C D_{3i}$  & $\C C_i$  \\
 $\qquad +|[(12)(45)]>$& & & \\
 & & & \\
$\{|[(12)(34)]>+|[(13)(25)]>+$  &$\tilde{D_5}$   &   $F(I/D_5)\tens \C D_{5i}$  & $\C C_{2i}$ \\
$+|[(15)(24)]>+ |[(23)(45)]>$             &                   &                               &           \\
$\qquad +|[(14)(35)]>\}$& & & \\
$|[123]>+|[132]>$   & & & \\
                        &$\tilde{C_3}$   &   $F(I/C_3)\tens \C D_{3i}$  & $\C C_{2i}$ \\
 & & & \\
$|[123]>+|[134]>+$  &$\Td$              &   $F(I/T)\tens \C T_i$       & $\C C_i$ \\
$\qquad  +|[142]>+|[243]>$& & & \\
& & & \\
$\{|[123]>+|[124]>+$ &$\Td$              &   $F(I/T)\tens \C T_i$       & $\C C_i$ \\
$\qquad +|[132]>+|[134]>+$ & & & \\
$+|[234]>+|[142]>+$        &                   &                               &       \\
$\qquad +|[143]>+|[243]>\}$ & & & \\
& & & \\
$\{|[123]>+|[152]>+$     &$\tilde{I}$     &   $\C C_{3i}$                 & $\C C_i$ \\
$\qquad +|[135]>+|[253]>+$ & & & \\
$+|[142]>+|[134]>+$&                   &                               &           \\
 $\qquad +|[243]>\}$& & & \\
 & & & \\
$\{|[12345]>+|[13524]>$                     & $\tilde{C_5}$  &   $F(I/C_5)\tens \C D_{5i}$  & $\C C_{2i}$\\
$+ |[14253]>+|[15432]>\}$                   &                   &                               &           \\
 & & & \\
\hline
\end{tabular}
\caption{{\small Combined defect condensates in an achiral
icosahedral nematic.}} \label{tab:icoscomb}
\end{table}
\subsubsection{Achiral icosahedral nematic}

The single defect condensates in $F(\Id)\tens\C I_i$ are given in
table \ref{tab:icosisd}. The class sum defect condensates are given
in table \ref{tab:icosclsum}. Finally, the combined defect
condensates are given in table \ref{tab:icoscomb}. The sample we
present in the table is basically exhaustive.
\subsubsection{Comments on the conjectured phases}

The tables given above, containing the possible phases induced by
defect condensation yield a lot of information of both physical and
mathematical interest. Yet, they don't tell the full story, as they
do not describe the hadronic composites which we alluded to earlier.
Information on these composites is laborious but straightforward to
extract, and we would like to comment on what kind of analysis would
reveal the hadronic content of a particular phase.

First we note that more often than not, the residual symmetry
algebra $\Rcal$ is not a Hopf algebra \footnote{One reads this off
the magnetic part of the Hopf symmetry. If it is of the form
$F(G/H)$, where $H$ is not a normal subgroup of $G$, then $G/H$ is
not a group, and one can prove that the algebra is not Hopf in this
case. If the magnetic part is of the form $F(H_m)$, where $H_m$ is a
group, then the residual symmetry algebra is a Hopf algebra.}. This
is perfectly acceptable, since there is no physical reason to assume
that $\Rcal$ is a Hopf algebra. The reason is subtle
\cite{bmbreaking:2006}: for a symmetry algebra to be a Hopf algebra,
it is necessary that the fusion of particles (i.e. the tensor
product of irreducible representations) be associative. Now some of
the irreducible representations of $\Rcal$ correspond to confined
excitations, which means that the condensate is in a different
internal state to the left and right of the excitation. Thus
particles to the left of this confined excitation ``see'' a
different condensate. This leads to the necessity of introducing an
ordering when taking the tensor product of representations, which
corresponds to specifying in which order the particles are brought
into the system.

The unconfined symmetry algebra $\Ucal$ corresponds to the
symmetry algebra whose irreducible representations are precisely
the unconfined representations of $\Rcal$. These unconfined
representations don't suffer from the necessity of introducing an
order, and therefore we expect $\Ucal$ to be a Hopf algebra. In
all the cases we've worked out this turns out to be the case.

From the tables we learn that different defect condensates may
induce the same $\Ucal$, however, they do lead to a different
$(\Rcal,\Ucal)$ pair. If two phases have the same $\Ucal$, their
low energy degrees of freedom share many properties (e.g. their
representation theory, their braiding properties). So to tell
these two phases apart, it may be necessary to probe unconfined
composites of confined excitations (which we called hadronic
composites). These may occur at a higher energy scale (depending
on the precise dynamics). The constituent structure of the
hadronic composites can be derived from $\Rcal$. To determine the
admissible composites, one must take tensor products of several
confined excitations and decompose the product into a direct sum
of irreducible representations. Every unconfined representation
that appears in such a decomposition, corresponds to a hadronic
composite. Note that even if $\Ucal=\C e$, the trivial Hopf
algebra, there can still be nontrivial composites, the
decomposition then has to yield the trivial representation (of
$\Rcal$). It is rather tedious to work out the sets of simplest
allowed composites, because the calculation is complicated by the
fact that $\Rcal$ is not a Hopf algebra, but it is
straightforward. We have refrained from carrying out such an
analysis at this stage.\\

\section{Conclusions}

In this paper we have applied the formalism for breaking quantum
double symmetries by defect condensates to some classes of rather
exotic non-Abelian nematics. We found a wide variety of
conceivable phases each characterized by a set of unconfined
degrees of freedom associated with an unconfined algebra $\Ucal$.
There may also be confined degrees of freedom described by an
intermediate symmetry algebra $\Tcal_r$. Clearly, whether such
phases will actually be realized in nature depends on the detailed
dynamics of these media. It would of course be of great value to
look for experimental parameters by which these phases could be
induced and furthermore to develop observable signatures by which
they could be distinguished. These important questions deserve
serious attention but are  beyond the kinematical scope of this paper.\\
{\bf Acknowledgement:} One of us (F.A.B.) likes to thank the
Yukawa Institute for Theoretical Physics of Kyoto University for
an inspiring visit during which part of this work was carried out.

\bibliographystyle{unsrt}
\bibliography{blob}

\end{document}